\documentclass{article}

\usepackage{arxiv}

\usepackage[utf8]{inputenc} 
\usepackage[T1]{fontenc}    
\usepackage[hypertexnames=false]{hyperref}      
\usepackage{url}            
\usepackage{booktabs}       
\usepackage{amsfonts}       
\usepackage{nicefrac}       
\usepackage{microtype}      
\usepackage{graphicx}
\usepackage{doi}
\usepackage[sort&compress,square,comma,authoryear]{natbib}
\usepackage{multirow}
\usepackage{tabularx}
\usepackage{xurl}
\usepackage{appendix}
\usepackage{makecell}
\usepackage{placeins}
\usepackage{fancyhdr}

\graphicspath{ {./images/} }

\title{AI in Debt Collection:\protect\\ Estimating the Psychological Impact on Consumers}

\author{ {Minou Goetze}\\
	Psychology School, Faculty of Business \& Media\\
	Fresenius University of Applied Sciences\\
	Hamburg, Germany\\
	\texttt{minou.goetze@hs-fresenius.de} \\
	\And
	{Sebastian Clajus} \\
	PAIR Finance GmbH\\
	Berlin, Germany\\
	\texttt{sebastian.clajus@pairfinance.com} \\
	\And
	{Stephan Stricker} \\
	PAIR Finance GmbH\\
	Berlin, Germany\\
	\texttt{stephan.stricker@pairfinance.com} \\
}

\date{January 19, 2026}

\renewcommand{\undertitle}{}

\begin{document}
\maketitle

\begin{abstract}
The present study investigates the psychological and behavioral implications of integrating AI into debt collection practices using data from eleven European countries. Drawing on a large-scale experimental design (n = 3514) comparing human versus AI-mediated communication, we examine effects on consumers' social preferences (fairness, trust, reciprocity, efficiency) and social emotions (stigma, empathy). Participants perceive human interactions as more fair and more likely to elicit reciprocity, while AI-mediated communication is viewed as more efficient; no differences emerge in trust. Human contact elicits greater empathy, but also stronger feelings of stigma. Exploratory analyses reveal notable variation between gender, age groups, and cultural contexts. In general, the findings suggest that AI-mediated communication can improve efficiency and reduce stigma without diminishing trust, but should be used carefully in situations that require high empathy or increased sensitivity to fairness. The study advances our understanding of how AI influences the psychological dynamics in sensitive financial interactions and informs the design of communication strategies that balance technological effectiveness with interpersonal awareness.
\end{abstract}


\newpage
\section{Introduction}
In an increasingly digital society, the intersection of advanced technologies and consumer interaction is fundamentally reshaping the service industries \citep{brynjolfssonmcafee14, zuboff19}. As organizations seek to remain competitive despite rapid technological and societal changes, they must critically examine whether their products, services, and communication strategies meet the expectations of an emerging generation of consumers—particularly younger, digitally native populations. These consumers prioritize emotional intelligence, transparency, and personalization in their interactions, especially when dealing with sensitive matters such as financial obligations \citep{goetzeetal23, tuongetal25}.

This transformation is especially salient in the domain of debt collection, where the stakes are high for both companies and consumers. Conventional, often rigid, and impersonal debt collection methods may not address the psychological and emotional needs of contemporary consumers, potentially undermining trust and cooperation \citep{brennanbaines08}. 

As a result, organizations are increasingly turning to Artificial Intelligence (AI) to enhance their customer engagement strategies. AI-driven solutions offer a range of potential advantages, such as greater personalization, context-aware communication, and improved adaptability-features that are particularly relevant in times of financial friction \citep{ghaffarietal21}.

This study investigates the psychological and behavioral implications of integrating AI into debt collection practices. Specifically, it examines 1) how AI-mediated communication influences consumers' social preferences regarding fairness, trust, reciprocity, and efficiency; and 2) how individuals emotionally respond to AI involvement in financial interactions, including feelings of stigma and perceptions of empathy.

To capture the variability of these phenomena in cultural and regulatory contexts, this study adopts a cross-national comparative design, collecting experimental data in eleven European countries: Germany, France, Portugal, Spain, Italy, the Netherlands, Sweden, Poland, Austria, Belgium, and Switzerland. Prior research has shown that consumer trust in technology, perceptions of fairness, and financial norms vary significantly between countries \citep{scantamburloetal24}, making a cross-national perspective essential. This approach will allow us to explore how attitudes towards AI in debt collection differ across national contexts and whether cultural, economic, or institutional factors shape consumer responses. By investigating psychological implications of AI-mediated communication, the current article offers empirical insights into the strategic integration of AI in debt collection and its broader implications for ethical consumer engagement, regulatory design, and financial communication in a rapidly digitizing Europe.

\section{Theoretical Background}
\label{sec:theoretical_background}

\subsection{Social Preferences: Fairness, Reciprocity, Trust}
AI-mediated communication introduces three psychologically relevant changes to ordinary interpersonal messaging. First, AI-mediated communication is a data-driven approach that does not take individual exceptions into consideration. This approach can undermine perceived fairness depending on the context. On a broader level, numerous studies have shown how preferences for fairness influence economic decision-making and judgment (e.g.,\citep{kahnemanetal86}. In addition, models of fairness and inequity aversion can explain behavior observed in many economic games, like the ultimatum game \citep{fehrschmidt99, fehrschmidt06}. Comparing fairness perceptions for humans and AI systems, previous literature suggests that humans are viewed as fairer compared to AI systems, in criminal justice \citep{harrisonetal20, wang18}, healthcare \citep{leerich21}, and work-related decisions \citep{acikgozetal20, bankinsetal22}, \citep{newmanetal20, nobleetal21}. Across a range of contexts, results on perceived fairness and AI systems are somewhat mixed, depending strongly on the respective context (for a review, see \citep{starkeetal22}). It is still unclear how these results extend to a financial context and if AI-mediated communication will be considered to be fair. 

Second, AI-mediated communication changes the type of interaction from a human-human interaction to a human-technology interaction. While this change might alter our evaluation of the interaction’s outcome (perceived as fair or unfair), it also raises the question of whether we read intentions into the technology’s action and, if so, how we respond to it. The social rule of reciprocity states that we return a favor or kindly respond to others’ generosity (positive reciprocity) and, on the other hand, retaliate hostile actions (negative reciprocity) \citep{rabin93, fehrgaechter00, bowlesgintis02, cialdini08}. Reciprocity plays a fundamental role within human-human interactions, such as employer-employee relationships \citep{kubeetal12}, cooperation in order to provide public goods \citep{fischbacheretal01} and enforce contracts \citep{fehretal97}, charitable giving \citep{falk07}, or even restaurant visits \citep{rindstrohmetz99}. Human agents are more readily attributed motives, emotions, and intentions, making them more likely than AI systems to elicit reciprocal behavior from consumers \citep{mendeetal24}. In contrast, AI agents, while efficient, may lack the perceived emotional depth necessary to foster reciprocal commitment. 

Third, the source of authorship is changed from a human to a technological device. The belief that a message was written by an algorithm or optimized by AI has previously shown to downgrade perceived trust in messages (e.g., \citep{daietal24}). Experimental evidence from trust games demonstrates that levels of trust and trustworthiness are important predictors for individuals’ cooperation \citep{bergetal95}, \citep{johnsonmislin11}. Neuroimaging research has identified specific neural mechanisms associated with trust decisions, including activation in brain regions linked to reward processing when trust is reciprocated and regions associated with negative emotions when trust is violated \citep{fehrcamerer07}. Field experiments have extended these laboratory findings to real-world settings, demonstrating that trust facilitates economic transactions in various contexts \citep{karlan05}, \citep{porteretal25}. Regarding consumer decisions in e-commerce, a recent study investigated the impact of AI-mediated communication on consumers’ trust and found that levels of trust are higher for humans than for AI-mediated communication \citep{hennighausenetal25}. Yet, it is unclear how levels of trust are affected when AI-mediated communication is used in financial communication. 

Based on previous findings, we assume that participants evaluating an AI-mediated conversation in a debt collection context will rate fairness lower than participants evaluating a human conversation (H1). Further, we hypothesize that participants evaluating an AI-mediated conversation will rate the content to be less trustful than participants evaluating a human conversation (H2). Finally, we hypothesize that after an AI-mediated conversation, participants will be less inclined to show reciprocal behaviors compared to a human conversation (H3).

\subsection{Social Emotions: Stigma and Empathy}
Historically, debt has carried a strong social stigma and continues to represent an emotionally charged and sensitive topic for many individuals \citep{andelicetal18}. As the use of AI becomes increasingly prevalent in sensitive service domains such as debt collection, understanding how consumers emotionally respond to AI versus human agents has become critical. Debt collection pressure functions as a relational stressor that undermines well-being by increasing financial strain, reinforcing feelings of shame and stigma, and triggering distress through negative interpersonal interactions with collectors \citep{rhodesetal25}. A recent study investigated consumers’ attitudes towards debt collection from a cross-generational perspective. The results revealed that debt collection is associated with negative stigma, even more so for older consumers \citep{goetzeetal23}. These findings underscore that debt collection is not merely a financial process but a deeply social and emotional encounter in which perceptions of the other party’s moral and empathic capacities matter.

A growing body of research indicates that people perceive human agents as possessing higher moral agency and greater capacity for social evaluation than AI or machines. For instance, \citep{bigmangray18} found that individuals resist machines making moral decisions because machines are perceived as lacking the human-like abilities to think, feel, or understand intent. Similarly, it was demonstrated that when evaluating moral transgressions, humans are judged as more intentional, blameworthy, and capable of moral reasoning than artificial agents, which are viewed as merely executing programmed instructions \citep{wilsonetal22}. Together, these findings suggest that people attribute richer moral cognition to humans than to AI. In emotionally charged contexts such as debt collection, this implies that consumers may experience stronger feelings of stigma and social evaluation when interacting with a human debt collector, who is perceived as capable of judging and morally evaluating them, than when engaging with an AI system that lacks such perceived moral capacity.

Emotionally sensitive interactions like debt collection also require empathy and emotional support to achieve constructive outcomes for both parties. Prior research comparing AI-mediated and human-mediated communication shows that responses believed to be human-authored are judged more empathetic: in a study comparing human- versus AI-written narratives in social support contexts, participants significantly empathized more with the stories attributed to humans than AI \citep{shenetal24}. Similarly, \citep{rubinetal25} demonstrated that when participants believed AI systems influenced an empathic message, perceived empathy ratings dropped sharply, suggesting that the belief about the source of a message critically shapes its emotional impact. These findings highlight that perceived humanness enhances emotional resonance and social connection in sensitive communication settings.

Based on these theoretical arguments and prior findings, we propose the following hypotheses: Perceived stigma will be higher for interactions with human assistants than interactions with AI assistants (H4). Empathy will be rated higher for interactions with human assistants than interactions with AI assistants (H5).

\section{Methodology}

\subsection{Participants}

Participants were recruited via online crowdsourcing platforms and eligible individuals were required to be between 18 and 70 years of age. We determined the target sample size of at least 376 participants by conducting an a priori power analysis (power = .95) based on an effect size estimate found in \citep{rubinetal25}\footnote{Cohen’s \textit{d} = 0.34 for the effect of human vs. AI on perceived empathy.}. A total of 3514 participants completed the study. The final sample had a mean age of 35 years (\textit{SD} = 11) and 48\% were female. Participants were drawn from several European countries, including Germany (\textit{n} = 1036), Austria (\textit{n} = 412), Switzerland (\textit{n} = 205), France (\textit{n} = 421), Belgium (\textit{n} = 60), the Netherlands \textit{n} = 647), Portugal (\textit{n} = 293), Italy (\textit{n} = 101), and Sweden (\textit{n} = 47). Participants completed the study in their respective native language. No prior experience with debt collection was required to partake in the study. All participants provided informed consent prior to participation and received compensation via the platform (average compensation was 1€, depending on the platform and country).

\subsection{Materials and Design}

The study employed a between-subjects experimental design. Participants were randomly assigned to one of two experimental treatments (Human assistant: \textit{n} = 1775, AI assistant: \textit{n} = 1739). In both treatments, participants read a scripted telephone conversation set in a debt collection scenario. In one treatment, the script featured an interaction with a human assistant, whereas in the other, the script featured an interaction with an AI assistant. Both scripts entailed a conversation between a debtor and an assistant that aimed at finding a solution for an unpaid invoice. The scripts can be found in the Appendix (Table~\ref{tab:script}). Based on previous research \citep{skantze21}, we identified the following key features that differentiated the script with an AI assistant from the script with a human assistant: 1) longer waiting time described for human interaction, 2) more pauses and less automatic processing for human interaction. The scripts were identical in terms of context (settling an open debt), the amount of the open debt, an offer to pay the debt in installments, the offered amount of monthly installments, and finding a successful solution with an installment plan.  

\subsubsection{Procedure}

Participants were recruited via the two online crowdsourcing platforms \textit{Clickworker} and \textit{Prolific} and received an invitation to participate in the study via e-mail. By clicking on the link, they were referred to the online survey platform Unipark (Tivian, Germany) where data was collected between August and October 2025. Participants were randomly assigned to one of two treatments. They started the survey by either reading the AI or human telephone script. Following the script, participants completed an online survey assessing their perceptions of the interaction and their attitudes toward the process. The outcome variables measured here included perceived fairness (e.g. “I felt that the process was fair and consistent”), trust in information provided (“I believe the information provided to me was accurate”), inclination to show reciprocal behavior (“If I were asked for a review after the call, I would be willing to write one”), efficiency (e.g. “My time during the call was used efficiently”), feeling of stigmatization (e.g. “I felt judged by my conversation partner”), and perceived empathy (“I felt understood and treated with empathy”).For a complete list of items that were used to measure these outcome variables, see Table~\ref{tab:items} in the appendix.\footnote{We also measured perceived autonomy, willingness to pay, and willingness to find a solution. See Table~\ref{tab:other_preferences} in the appendix for the respective results.} Participants indicated their assessment on a 5-point Likert scale, ranging from 1 = strongly disagree to 5 = strongly agree. Afterward, they responded to the 10-item Big Five Inventory (BFI-10) to capture personality traits across the five dimensions (openness, conscientiousness, extraversion, agreeableness, and neuroticism). Finally, participants provided demographic information including age, gender, country of residence, and education. Completing the survey took participants around 5-10 minutes. 

\subsection{Data Analysis}

To examine the effects of assistant type (human assistant vs. AI assistant) on participants’ perceptions and evaluations, we will estimate a series of ordered logistic regression models. Since the outcome variables of interest are ordinal rather than continuous, we selected the ordered logit model to analyze our data. For each outcome, we will fit a separate ordered logit regression model in which the primary predictor is assistant type (coded as 1 = AI assistant, 0 = human assistant). To account for possible demographic influences, we include age (continuous) and gender (binary) as control variables in all models. All models controlled for country-specific differences using fixed effects to account for potential confounding (reference = Germany). To examine whether the effect of communication type (human vs. AI) varied across individual characteristics, we included interaction terms between treatment and participants’ age and gender in an ordered logistic regression model. All regression analyses were estimated using robust standard errors to account for potential heteroskedasticity in the error terms. Because the number of countries in the sample (N = 11) was relatively small, standard errors were not clustered at the country level. Results will be reported as coefficients and robust standard errors. We used a significance level of $\alpha = 0.05$ when testing all hypotheses. 

\section{Results}

\subsection{Social Preferences and Efficiency}
First, we analyzed the effects of treatment differences between communication with a human assistant and an AI assistant on participants’ social preferences. Descriptive results show that for communication with a human assistant, levels of perceived fairness (\textit{M} = 4.33, \textit{SD} = 0.78) were higher than for an AI assistant (\textit{M} = 4.24, \textit{SD} = 0.77). Consistent with this pattern, a slightly larger share of participants in the human treatment (51\%) than in the AI treatment (49\%) rated the process as fair. Similarly, intentions to reciprocate were marginally higher for human assistants (\textit{M} = 3.91, \textit{SD} = 1.07) than for AI assistants (\textit{M} = 3.85, \textit{SD} = 1.06), with 52\% of participants in the human treatment and 48\% in the AI treatment reporting intentions to reciprocate.

\begin{figure}[ht]
	\centering
	\caption{Treatment effects displaying AI vs. human for social preferences and emotions. Error bars indicate 95\% confidence intervals.}
	\includegraphics[width=16cm, height=12cm]{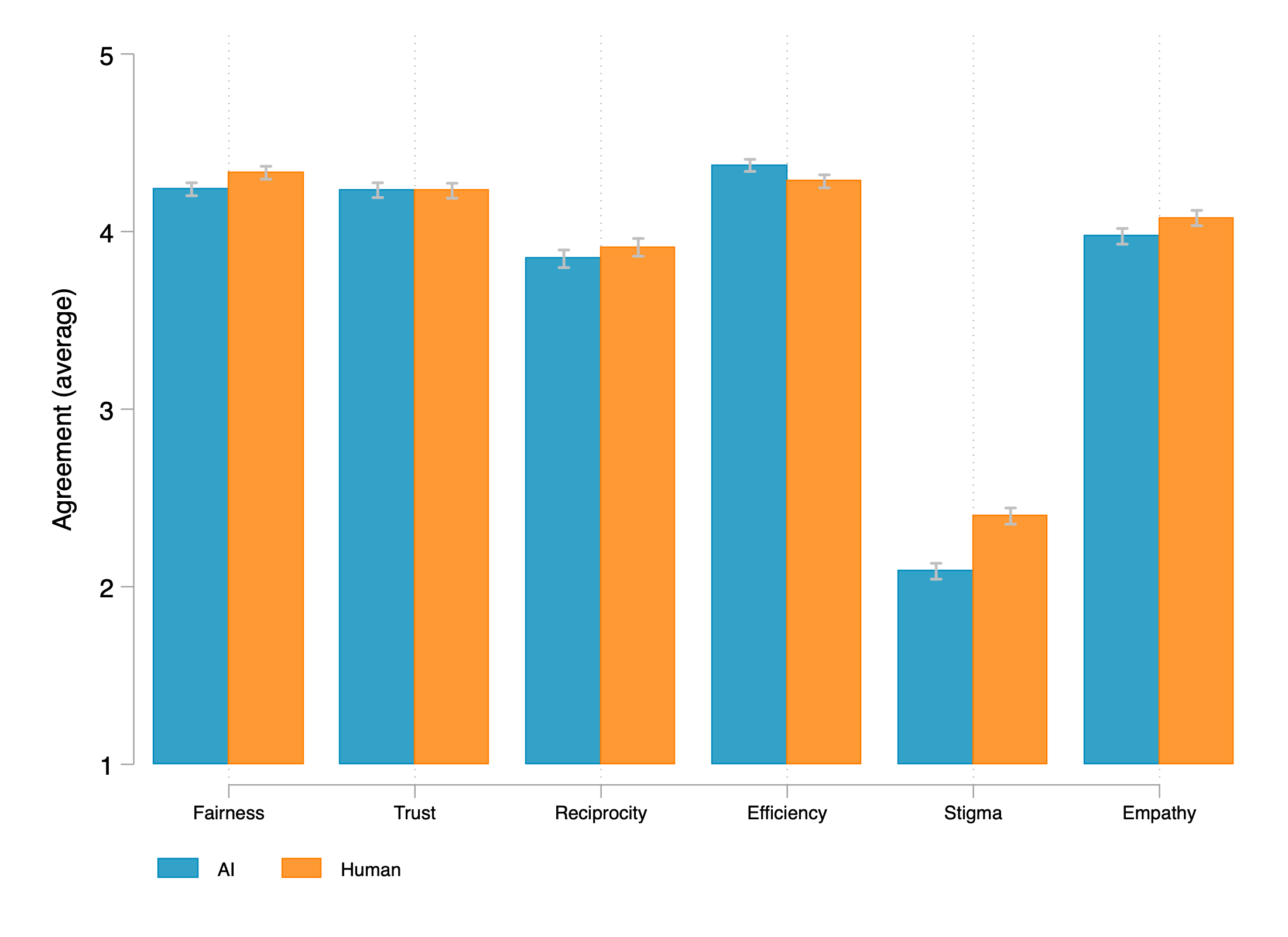}
	\label{fig:treatment}
\end{figure}

In contrast, trust in the information provided did not differ between conditions, both in terms of mean ratings (human: \textit{M} = 4.23, \textit{SD} = 0.91; AI: \textit{M} = 4.23, \textit{SD} = 0.89) and the share of participants reporting high trust (50\% in both treatments). Finally, perceived efficiency of the interaction was rated slightly higher for communication with an AI assistant (\textit{M} = 4.37, \textit{SD} = 0.72) than with a human assistant (\textit{M} = 4.28, \textit{SD} = 0.79), with the proportion of participants reporting high efficiency being slightly higher in the AI treatment (human: 49\%; AI: 50\%). For an overview of treatment differences between AI and human assistants, see Figure~\ref{fig:treatment}.

To test for statistically significant results, we conducted ordered logit regressions, controlling for age, gender, and country. The results indicate that levels of perceived fairness and intentions to reciprocate were significantly higher for human assistants than for AI assistants. Based on the model, the predicted probability of rating the process as fair was 87\% in the AI treatment and 90\% in the human treatment, while the predicted probability of reporting intentions to reciprocate was 69\% for AI and 72\% for human assistants. These results support our H1 and H3. On the other hand, efficiency was rated significantly higher for AI assistants than for human assistants, with predicted probabilities of 91\% for AI and 89\% for human assistants. Regarding levels of trust in presented information, in contrast to H2, we did not find any treatment effect; the predicted probabilities were similar across treatments (84\% for AI and 85\% for human assistants). For a complete overview of the regression results, see For a complete list of items that were used to measure these outcome variables, see Table~\ref{tab:regression_preferences}. We found considerable differences regarding the influence of age and gender on social preferences in this context. Specifically, older participants were more likely to consider the interaction as fair, more likely to trust in information presented to them, more inclined to reciprocate behavior, and more likely to rate the process as efficient. Similarly, female participants were significantly more likely to consider the interaction as fair, trust in information presented to them, reciprocate behavior, and rate the process as efficient.  

\begin{table}[ht]
	\caption{Predicting social preference rating using assistant type, age, gender, and country of residence via ordered logistic regression.}
	\centering
    \begin{tabular}{lcccccc}
        \toprule
         & Fairness & Trust & Reciprocity & Efficiency & Stigma & Empathy \\
        \midrule
        AI assistant (Human=1, AI=0)
            & 0.278*** & 0.005 & 0.155* & -0.204** & 0.606*** & 0.234*** \\
            & (0.062) & (0.065) & (0.063) & (0.062) & (0.060) & (0.063) \\
        Age 
            & 0.007* & 0.010*** & 0.015*** & 0.006 & -0.001 & 0.008** \\
            & (0.003) & (0.003) & (0.003) & (0.003) & (0.003) & (0.003) \\
        Gender female
            & 0.370*** & 0.381*** & 0.505*** & 0.403*** & -0.192** & 0.495*** \\
            & (0.062) & (0.065) & (0.063) & (0.062) & (0.060) & (0.064) \\
        \midrule
        Observations 
            & 3492 & 3492 & 3492 & 3492 & 3492 & 3492 \\
        \bottomrule
    \end{tabular}
    \small
    \raggedright
    \break
    \textit{Note.} All models include country fixed effects (coefficients not shown for clarity). Full coefficients for country fixed effects are available in the Appendix Table~\ref{tab:full_country}. Ordered logit coefficients are reported. Robust standard errors in parentheses. * \textit{p} < .05, ** \textit{p} < .01, *** \textit{p} < .001.
	\label{tab:regression_preferences}
\end{table}

\subsection{Social Emotions}

Next, we analyzed the effects of treatment differences between communication with a human assistant and an AI assistant on participants’ social emotions. Descriptive results show that levels of perceived stigma were higher for communication with a human assistant (\textit{M} = 2.40, \textit{SD} = 0.98) as compared to communication with an AI assistant (\textit{M} = 2.09, \textit{SD} = 0.95). Consistent with this pattern, a larger proportion of participants in the human treatment (61\%) than in the AI treatment (39\%) reported feeling stigmatized. Regarding perceived empathy, the results indicate that perceived empathy was higher for communication with a human assistant (\textit{M} = 4.08, \textit{SD} = 0.93) than for communication with an AI assistant (\textit{M} = 3.97, \textit{SD} = 0.94), with 53\% of participants in the human treatment and 47\% in the AI treatment reporting high levels of empathy.

Testing for statistical significance, the results of the ordered logit models indicate that both effects for social emotions are statistically significant (see Table~\ref{tab:regression_preferences}). Specifically, participants reported stronger feelings of stigma when interacting with a human assistant compared to an AI assistant. Based on the model, the predicted probability of feeling stigmatized was 11\% in the AI treatment and 19\% in the human treatment. At the same time, participants perceived higher levels of empathy from the human assistant than from the AI assistant, with predicted probabilities of 75\% for AI and 79\% for human assistants. These findings support H4 and H5. 

\begin{figure}[ht]
	\centering
	\caption{ Interaction effect displaying increase of difference between AI vs. human for perceived stigma across age. Error bars indicate 95\% confidence intervals.}
	\includegraphics[width=16cm, height=12cm]{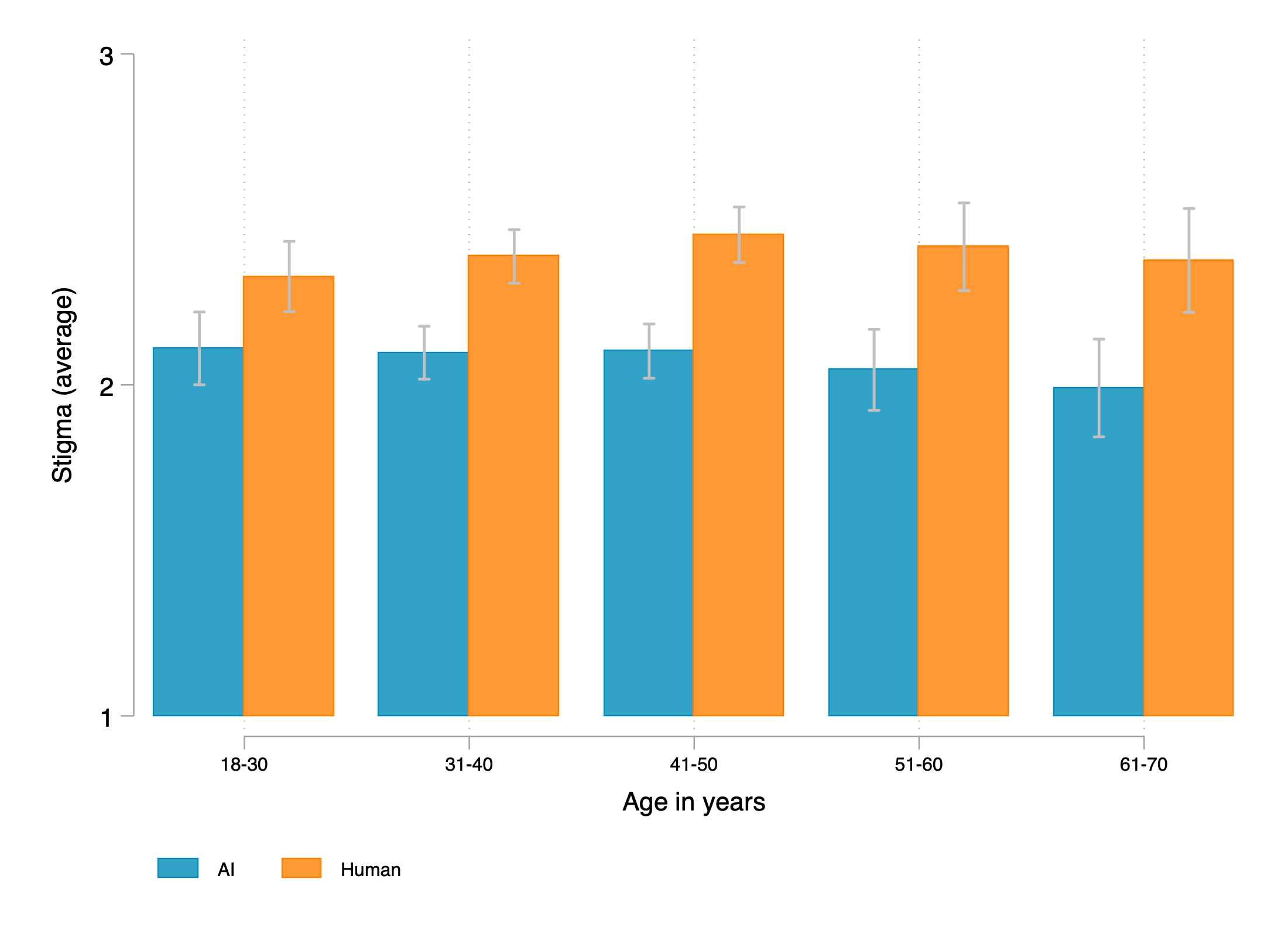}
	\label{fig:interaction}
\end{figure}

In line with our results on social preferences, we also found notable effects for age and gender on social emotions. Specifically, older participants were more likely to rate their interaction partner as empathic while there were no significant effects regarding stigma. Female participants were more likely to rate their interaction partner as empathic. Interestingly, male participants were significantly more likely to feel stigmatized by their interaction partner than female participants (see Table~\ref{tab:regression_preferences}). Going one step further, we tested whether the treatment effects differed depending on participants’ age and gender. A significant interaction between treatment condition and age (\textit{b} = 0.01, \textit{SE} = 0.01, \textit{p} = .012) indicated that the effect of assistant type on perceived stigma became stronger with age (see Figure~\ref{fig:interaction}). In other words, younger participants showed relatively small differences in stigma across assistant types, whereas older participants perceived notably more stigma in interactions with human assistants compared to AI assistants. We did not find significant interaction effects between assistant type and gender (see full results on interaction effects in Table~\ref{tab:interaction_preferences}).

\FloatBarrier
\subsection{Explorative Analysis of Country-specific Differences}
To examine whether the effect of communication type (human vs. AI) varied across countries, we estimated an ordered logistic regression model predicting each social preference and social emotion as a function of treatment (human vs. AI), country of residence, and their interaction. Age and gender were included as covariates. We excluded countries with less than 100 observations from the analysis, such as Belgium (\textit{n} = 60), Sweden (\textit{n} = 47) and countries outside of Europe (\textit{n} = 8). See Table~\ref{tab:descriptives} in the appendix for an overview of descriptive statistics for all social preferences and emotions.

\begin{figure}[ht]
	\centering
	\caption{Differences in social preference ratings depending on treatment across countries for (a) fairness, (b) trust, (c) reciprocity, and (d) efficiency. Average agreement is presented on the y-axis, ranging from 1 (strongly disagree) to 5 (strongly agree). Error bars indicate 95\% confidence intervals.}
	\includegraphics[width=16cm, height=12cm]{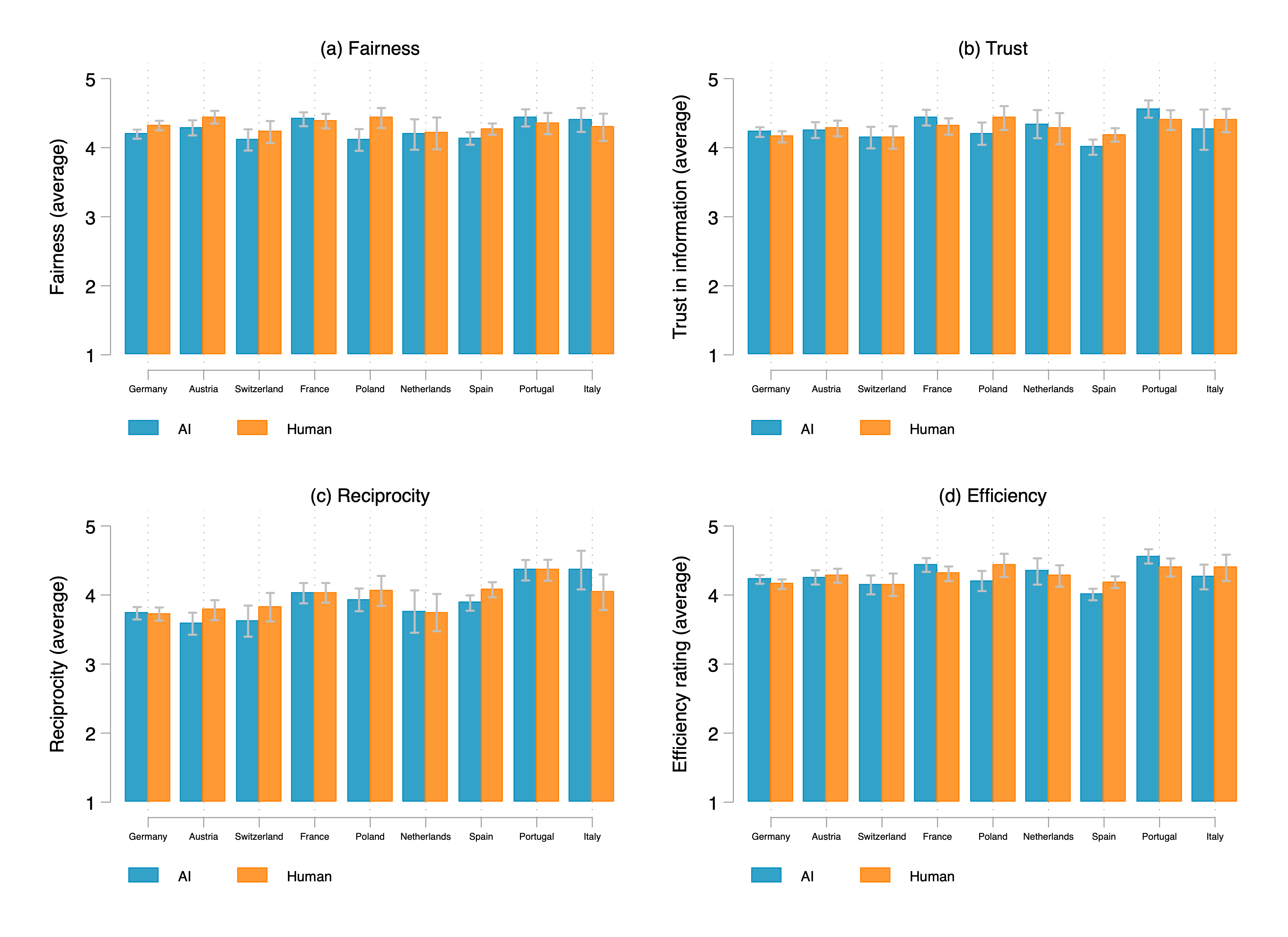}
	\label{fig:social_preferences}
\end{figure}

Analyzing the link between country and social preferences, we find substantial differences across countries (see Figure~\ref{fig:social_preferences} for an overview of average social preference ratings across all countries). Perceived fairness differed across countries, with significantly higher ratings in France (\textit{b} = 0.61, \textit{SE} = 0.15, \textit{p} < .001) and Portugal (\textit{b} = 0.77, \textit{SE} = 0.18, \textit{p} < .001) compared to Germany. No significant interaction between treatment and country emerged, suggesting that the effect of communication type on fairness was consistent across countries. Trust levels varied across countries, being lower in Spain (\textit{b} = -0.39, \textit{SE} = 0.13, \textit{p} = .004 ) and higher in France (\textit{b} = 0.66, \textit{SE} = 0.17, \textit{p} < .001) and Portugal (\textit{b} = 1.06, \textit{SE} = 0.20, \textit{p}  < .001) than in Germany. In Spain and Poland, the difference between trust in human and AI communication was significantly larger than in other countries, with participants expressing notably greater trust in human interactions (Spain: \textit{b} = 0.48, \textit{SE} = 0.20, \textit{p} = .012; Poland: \textit{b} = 0.68, \textit{SE} = 0.31, \textit{p} = .027). 

Intentions to reciprocate were higher in France (\textit{b} = 0.66, \textit{SE} = 0.16, \textit{p} < .001), Portugal (\textit{b} = 1.40, \textit{SE} = 0.19, \textit{p} < .001), and Italy (\textit{b} = 1.27, \textit{SE} = 0.29, \textit{p} < .001) than in Germany. As with trust, a significant interaction emerged for Spain, where the difference between human and AI communication was particularly pronounced (\textit{b} = 0.44, \textit{SE} = 0.18, \textit{p} = .017), meaning that participants were more willing to reciprocate after human contact. Efficiency ratings were generally higher in France (\textit{b} = 0.86, \textit{SE} = 0.16, \textit{p} < .001), Portugal (\textit{b} = 1.21, \textit{SE} = 0.19, \textit{p} < .001), and Italy (\textit{b} = 0.58, \textit{SE} = 0.28, \textit{p} = .037) than in Germany. Participants in France (\textit{b} = -0.58, \textit{SE} = 0.22, p = .008) and Portugal (\textit{b} = -0.68, \textit{SE} = 0.26, \textit{p} = .009) rated AI communication as particularly efficient, whereas in Poland efficiency was rated higher for humans (\textit{b} = 0.60, \textit{SE} = 0.29, \textit{p} = .041). 

\begin{figure}[ht]
	\centering
	\caption{Differences in social emotion ratings depending on treatment across countries for (a) stigma and (b) empathy. Average agreement is presented on the y-axis, ranging from 1 (strongly disagree) to 5 (strongly agree). Error bars indicate 95\% confidence intervals.}
	\includegraphics[width=16cm, height=6cm]{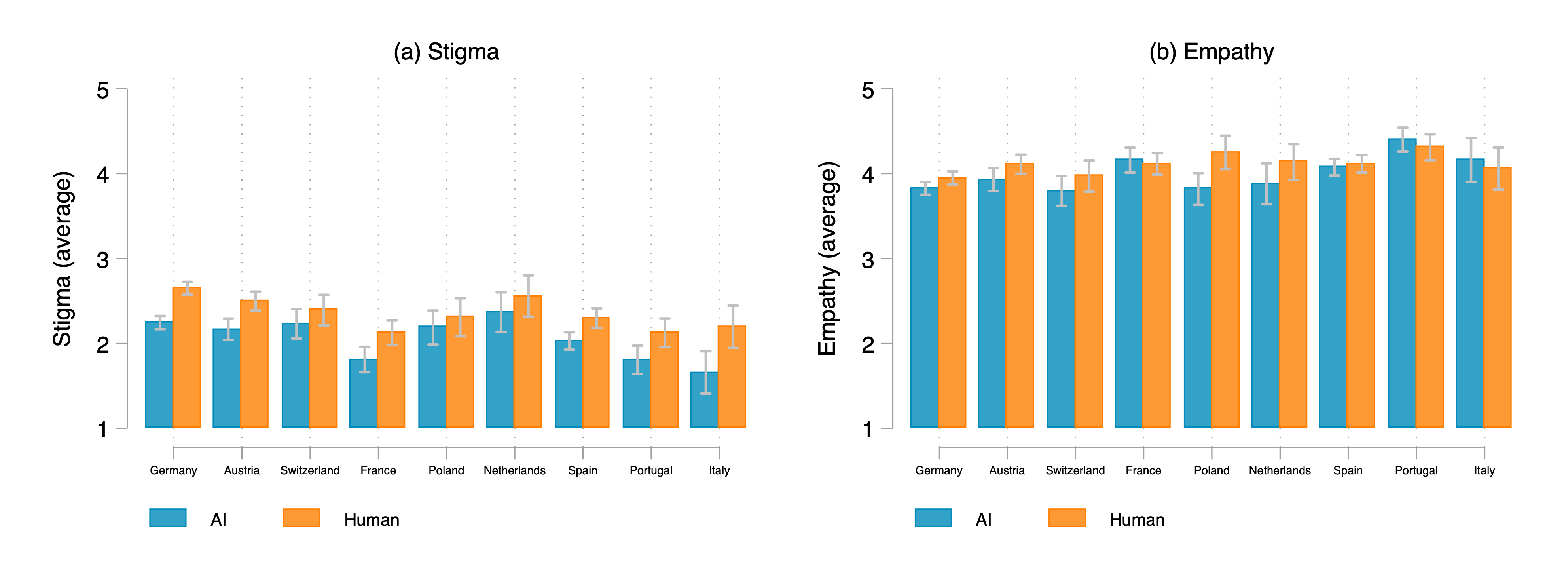}
	\label{fig:social_emotions}
\end{figure}

Next, we investigated the link between country and social emotions and again found considerable differences across countries (see Figure~\ref{fig:social_emotions} for an overview of average social emotion ratings across all countries). Perceived stigma was rated lower in France (\textit{b} = -1.13, \textit{SE} = 0.16, \textit{p} < .001), Spain (\textit{b} = -0.43, \textit{SE} = 0.13, \textit{p} = .001), Portugal (\textit{b} = -1.12, \textit{SE} = 0.18, \textit{p} < .001), and Italy (\textit{b} = -1.27, \textit{SE} = 0.27, \textit{p} < .001) than in Germany. Differences between AI and human communication were largest in Germany with higher stigma rating for human communication and smallest in Poland. Yet, differences between AI and human communication were not significantly different across countries. Empathy ratings were higher in France (\textit{b} = 0.99, \textit{SE} = 0.16, \textit{p} < .001), Spain (\textit{b} = 0.54, \textit{SE} = 0.13, \textit{p} < .001), and Portugal (\textit{b} = 1.53, \textit{SE} = 0.19, \textit{p} < .001) compared to Germany. The treatment effect differed by country: in France (\textit{b} = -0.49, \textit{SE} = 0.22, \textit{p} = .025) and Portugal (\textit{b} = -0.52, \textit{SE} = 0.26, \textit{p} = .045), empathy ratings did not significantly vary between AI and human communication, while in Poland humans were perceived as much more empathic than AI (\textit{b} = 0.66, \textit{SE} = 0.30, \textit{p} = .030). See Table~\ref{tab:interaction_country} for the complete regression results.

\FloatBarrier
\section{Discussion}
The present study provides novel insights into the psychological consequences of integrating AI into consumer service communication, revealing a pattern of social and emotional responses. Regarding social preferences, participants perceived human-mediated interactions as fairer and elicit stronger reciprocal intentions compared to AI-mediated interactions, whereas AI-mediated communication was evaluated as more efficient. Interestingly, no significant difference emerged for trust in information. Emotionally, human interactions were associated with higher perceived empathy but also with heightened feelings of stigma relative to AI interactions. These findings suggest that AI-mediated service communication offers notable advantages, including greater procedural efficiency and reduced feelings of stigma, without undermining trust in the information provided. Yet, AI systems should be deployed cautiously in situations that require high levels of empathy or where concerns about fairness and reciprocity are likely to shape consumer responses.

These findings extend prior research showing that algorithmic decision-making can reduce perceived procedural fairness and legitimacy \citep{araujoetal20, longonietal19}, and that AI agents are often seen as less capable of conveying authentic empathy \citep{perry23,dorigonigiardino25}. Consistent with the present findings, earlier studies have demonstrated that the “algorithmic gap” in perceived warmth and fairness can attenuate cooperation and prosocial responses in technology-mediated exchanges. At the same time, the present results are in line with previous research showing higher perceived efficiency for AI decision-making \citep{yuli22}. Findings regarding reciprocity are in line with previous studies indicating that humans are less likely to reciprocate when interacting with a machine \citep{vonschenketal25}. While levels of reciprocity might be lower for machines than humans, the present study suggests that reciprocity plays a role even when interacting with AI, which is supported by another study on human-robot interactions \citep{mobergetal24}.

Conversely, the higher efficiency evaluations for AI align with evidence that algorithmic systems are valued for consistency and speed in service delivery \citep{gursoyetal19}. The feeling of greater stigma in the human treatment may reflect the heightened salience of moral accountability and interpersonal evaluation when communication involves another person \citep{tangneyetal07}, leading to stronger self-conscious emotional reactions such as embarrassment or shame.

In an explorative analysis, we investigated national differences regarding attitudes towards AI across nine European countries. Taken together, the results reveal consistent regional patterns. Southern European countries (France, Portugal, Spain, and Italy) tended to report higher levels of fairness, reciprocity, and empathy, as well as lower stigma, compared to Germany. However, sensitivity to AI versus human communication varied: while participants in France and Portugal attributed greater efficiency to AI, those in Spain and Poland were more responsive to human communication in terms of trust, reciprocity, and empathy. This highlights that cultural context shapes both social and emotional responses to AI-mediated service communication. Our findings resonate with previous research showing that consumer responses to AI, and technology in general, are mediated by cultural and institutional norms \citep{gillespieetal25,hofstede01}. For example, evidence suggests notable differences between countries with advanced and emerging economies: People in emerging economies reported greater trust, acceptance and adoption of AI, higher levels of AI literacy and training, and more realized benefits from AI \citep{gillespieetal25}. A recent study investigated attitudes towards AI technologies across Europe and their results show a more positive attitude in Italy, Poland, Romania, and Spain, whereas there was lower trust in France, Germany, Netherland, and Sweden \citep{scantamburloetal24}. On a global level, there are even more pronounced geographic differences in the opinion about AI with industrial (including European) countries exhibiting a less positive attitude compared to Asian or African countries \citep{loewenetal24}.

Our findings demonstrate that individual differences play an important role: older and female participants reported higher perceptions of fairness, trust, reciprocity, efficiency, and empathy, while male participants reported greater feelings of stigma. Moreover, the influence of assistant type on perceived stigma increased with age, such that older participants felt more stigmatized by humans than by AI assistants. These patterns complement prior research showing that age and gender shape how people interpret and trust socio-technical communications and empathic signals (e.g., \citep{mendeetal24}), and extend evidence that individual characteristics moderate responses to service communications. In line with previous studies (e.g., \citep{goetzeetal23}), our results argue against “one-size-fits-all” communication approaches and support the need for adaptive, individual-level tailoring of service communications.

Beyond the debt collection context, these results have important implications for understanding how AI-mediated communication reshapes the psychological dynamics of economically and morally sensitive interactions. As financial institutions, service providers, and public-sector agencies increasingly employ AI interfaces for tasks involving compliance, negotiation, or repayment, the balance between efficiency and relational sensitivity becomes critical. The findings underscore that automation not only modifies communication channels but also reconfigures the emotional reactions to these interactions. Theoretically, this contributes to emerging work on the social cognition of human–AI relationships by illustrating that AI systems can influence perceptions of fairness, empathy, and self-worth, core drivers of cooperation and trust in socio-economic contexts. Practically, organizations may need to design hybrid approaches that retain human oversight or human-like cues in AI-mediated financial communication to mitigate potential costs. More broadly, the study points to the necessity of embedding psychological considerations into the deployment of AI systems across domains where social and emotional sensitivities are integral to consumer experience and societal trust.

There are various potential paths for future research. With regards to social preferences, research could examine how perceptions of fairness, trust, and reciprocity evolve over time with repeated interactions. Do the observed differences between human and AI assistants diminish as users become more familiar with AI systems? Research could also investigate possibilities to enhance reciprocity in AI interactions (e.g. communication styles) to bridge the reciprocity gap between human and AI assistants. While we found no difference in trust between human and AI assistants overall, future research might explore whether this equivalence holds across different domains (e.g., financial advice, health information, creative collaboration). Given the complementary strengths we identified (humans rated higher on fairness and reciprocity while AI rated higher on efficiency), following research could explore the above-mentioned hybrid interaction models that leverage both human and AI capabilities to optimize social outcomes. As for social emotions, future research could inform the design of sensitive AI services in domains where stigma is a barrier to help-seeking (e.g. mental health support or financial counseling for debt). While AI assistants were rated lower on empathy in our study, research should investigate approaches to enhance perceived empathy in AI interactions (e.g., personalization techniques) that could ultimately bridge the empathy gap between human and AI assistants.

\section{Conclusion}

In conclusion, the overall results of the study indicate that AI-mediated service communication can enhance procedural efficiency and reduce feelings of stigma without compromising trust in the information provided. The presented evidence supports the notion that, despite these advantages, AI systems should be applied cautiously in contexts that demand high empathy or heightened sensitivity to fairness and reciprocity. Moreover, the findings demonstrate that responses to service communication vary meaningfully across gender, age groups and cultural contexts, underscoring the need to move beyond uniform communication strategies. Together, these insights highlight the need for adaptive, individually tailored communication strategies when integrating AI into service environments.

\newpage

\bibliographystyle{plainnat}


\newpage

\section*{Appendix}
\setcounter{table}{0}
\renewcommand{\thetable}{A\arabic{table}}

\begin{table}[ht]
    \caption{Telephone script for human and AI treatment.}
    \centering
    \small 
    \begin{tabular}{>{\raggedright\arraybackslash}p{2cm} >{\raggedright\arraybackslash}p{12cm}}
        \toprule
        Assistant type & Content of script \\
        \midrule
        Human & Please try to put yourself in the following situation: \newline
        
        You ordered a pair of Bluetooth headphones, model SoundPro X300, from an online retailer. The purchase price was €89.99. You chose to pay by invoice, as this seemed financially manageable at the time. A few weeks later, your financial situation unexpectedly worsened due to a high utility bill you had to pay. The online retailer sent you a payment reminder via email. You ignored the reminder because you were unable to pay the invoice at that point. A debt collection agency was then commissioned by the retailer and has now contacted you. \newline
        
        Note: The conversation takes place over the phone with a representative of the company. You will make the call between 9 a.m. and 6 p.m., as those are the service hours. Before the conversation begins, you’ll be on hold for 10 minutes. \newline
        
        Below is a script of the phone call: \newline
        
        You: Good day. I received a letter from you today regarding an outstanding debt. I wanted to get in touch directly, as I’m currently unable to pay the full amount at once.\newline
        
        Representative: Good day. Please provide me with your reference number. [Pause while reference number is given]
        Thank you. This concerns an unpaid amount of €89.99. The invoice has been overdue for some time and we haven’t received any payment yet. I’d like to discuss with you how we can resolve this together. When do you think a payment might be possible?\newline
        
        You: Unfortunately, I don’t have the amount at the moment. After receiving an unexpectedly high utility bill, I’m financially overwhelmed. I’m sorry I didn’t respond to the reminder — I was simply too stressed.\newline
        
        Representative: I see… that does sound like a difficult situation. But of course, it would have been better if you had contacted us earlier.\newline
        
        You: I just wasn’t able to do that earlier, unfortunately.\newline
        
        Representative: Alright, let’s look ahead. So you're saying a one-time payment isn’t possible right now, correct?\newline
        
        You: That’s right. If possible, I’d prefer to pay in installments.\newline
        
        Representative: That’s generally possible. I’ll need to check something quickly… [Pause]
        So, how much could you afford to pay per month?\newline
        
        You: I’m not entirely sure, let me think for a moment. I’d say somewhere between €15 and €20 would be doable.\newline
        
        Representative: Okay, if we spread the amount over 4 months, that would be about €20 per month. That could work — how does that sound?\newline
        
        You: That sounds doable. I’m very grateful we can arrange it this way.\newline
        
        Representative: I’m glad to hear that. I’ll send you the agreement by post and email. If you run into any issues with the payments, please let us know early — we can adjust if needed. I wish you all the best.\newline
        
        You: Thank you for your help. \\[1em]
       \bottomrule
    \end{tabular}
    \label{tab:script}
\end{table}

\small 
\begin{table}[ht]
    \begin{tabular}{>{\raggedright\arraybackslash}p{2cm} >{\raggedright\arraybackslash}p{12cm}}
        \toprule
        Assistant type & Content of script \\
        \midrule
        AI & Please try to put yourself in the following situation: \newline

        You ordered a pair of Bluetooth headphones, model SoundPro X300, from an online retailer. The purchase price was €89.99. You chose to pay by invoice, as this seemed financially manageable at the time. A few weeks later, your financial situation unexpectedly worsened due to a high utility bill you had to pay. The online retailer sent you a payment reminder via email. You ignored the reminder because you were unable to pay the invoice at that point. A debt collection agency was then commissioned by the retailer and has now contacted you.\newline
        
        Note: The conversation takes place over the phone with an automated voice assistant (based on artificial intelligence). You can make the call at any time, as the service is available 24/7. There is a 10-second hold before the conversation begins.\newline
        
        Below is a script of the phone call:\newline
        
        You: Good day. I received a letter from you today regarding an outstanding debt. I wanted to get in touch directly because I’m currently having difficulty paying the full amount at once.\newline
        
        Digital Assistant: Thank you for calling. I can see your file here. The outstanding amount is €89.99 and has been overdue for some time. We’d like to find a solution together with you. Can you tell us if and when a payment would be possible?\newline
        
        You: Unfortunately, I don’t have the funds at the moment. I’m financially overwhelmed after receiving an unexpectedly high utility bill. I’m sorry I didn’t respond to the reminder — I was just really stressed.\newline
        
        Digital Assistant: Thank you for being honest. Our goal is to find a solution together with you. Would you be able to pay the debt in installments?\newline
        
        You: An installment plan would help. What would that look like?\newline
        
        Digital Assistant:
        There are different options. Based on your information, I’ve prepared the following proposal:\newline
        Installment: €20, Start date: 01/08/2025, Final installment: 01/11/2025 (includes a small remaining balance)\newline
        Would you like to accept this offer?\newline
        
        You: Yes, that sounds good to me.\newline
        
        Digital Assistant: Alright, thank you for confirming. You’ll receive all the details via email and post. If you run into financial difficulties again, please contact us early. We’ll work together to adjust the installments. I wish you all the best!\newline
        
        You: Thank you for your help. \\
        \bottomrule
    \end{tabular}
    \label{tab:script2}
\end{table}

\begin{table}[ht]
    \caption{List of items.}
    \centering
    \begin{tabular}{>{\raggedright\arraybackslash}p{1cm} >{\raggedright\arraybackslash}p{11cm}>{\raggedright\arraybackslash}p{2.5cm}}
        \toprule
        Number & Item & Concept \\
        \midrule
        1 & I felt I had full control over the repayment process of the outstanding invoice. & Autonomy \\
        2 & I felt that I was free to choose a payment solution. & Autonomy \\
        3 & I felt that the process was fair and consistent. & Fairness \\
        4 & I felt that the outcome was fair considering my situation. & Fairness \\
        5 & I would comply with the installment agreement after this conversation. & Payment \\
        6 & I would repay the full amount after this conversation. & Payment \\
        7 & I felt judged by my conversation partner. & Stigma \\
        8 & I felt uncomfortable having the phone call. & Stigma \\
        9 & I felt understood and treated with empathy. & Empathy \\
        10 & My time during the call was used efficiently. & Efficiency \\
        11 & We found a solution to the problem very quickly. & Efficiency \\
        12 & I believe the information provided to me was accurate. & Trust \\
        13 & If I were asked for a review after the call, I would be willing to write one. & Reciprocity \\
        14 & If I were in the same situation again and contacted by email, I would prefer a response from an AI (instant, but possibly generic) rather than from a human (slower, but possibly more specific). & Preference for AI \\
        15 & If I were in the same situation again, this phone call would encourage me to seek a solution earlier. & Solution \\
        \bottomrule
    \end{tabular}
    \label{tab:items}
\end{table}

\setcounter{figure}{0}
\renewcommand{\thefigure}{A\arabic{figure}}

\FloatBarrier

\begin{table}[ht]
	\caption{Predicting Autonomy, Willingness to find a solution, and Willingness to pay using assistant type, age, gender, and country of residence via ordered logistic regression.}
	\centering
	\begin{tabular}{lccc}
		\toprule
		   & Autonomy & Willingness to find Solution & Willingness to Pay \\
		\midrule
        Assistant type (Human=1, AI=0) & 0.192*** & 0.310*** & 0.014 \\
          & (0.060) & (0.064) & (0.060) \\
        Age & -0.000 & 0.012*** & -0.001* \\
          & (0.003) & (0.003) & (0.060) \\
        Gender female & 0.221*** & 0.425*** & 0.042 \\
          & (0.060) & (0.063) & (0.060) \\
        \bottomrule
        Observations & 3492 & 3492 & 3492 \\
		\bottomrule
	\end{tabular}
        \small
    \raggedright
    \break
    \textit{Note.} All models include country fixed effects (coefficients not shown for clarity). Ordered logit coefficients are reported. Robust standard errors in parentheses. * \textit{p} < .05, ** \textit{p} < .01, *** \textit{p} < .001. \\
	\label{tab:other_preferences}
\end{table}

\begin{table}[ht]
	\caption{Predicting social preference rating using assistant type, age, gender, and country of residence via ordered logistic regression, including full coefficients for country fixed effects.}
	\centering
	\begin{tabular}{lcccccc}
		\toprule
		   & Fairness & Trust & Reciprocity & Efficiency & Stigma & Empathy \\
		\midrule
        Assistant type (Human = 1, AI = 0) & 0.275*** & 0.004 & 0.154* & -0.205*** & 0.608*** & 0.233*** \\
        & (0.062) & (0.065) & (0.063) & (0.062) & (0.060) & (0.063) \\
        Age & 0.007* & 0.010*** & 0.015*** & 0.006* & -0.002 & 0.008** \\
        & (0.003) & (0.003) & (0.003) & (0.003) & (0.003) & (0.003) \\
        Gender female & 0.370*** & 0.382*** & 0.505*** & 0.404*** & -0.191** & 0.495*** \\
        & (0.062) & (0.065) & (0.063) & (0.062) & (0.060) & (0.064) \\
        Germany & 0.000 & 0.000 & 0.000 & 0.000 & 0.000 & 0.000 \\
        & (.) & (.) & (.) & (.) & (.) & (.) \\
        Austria & 0.348** & 0.221* & 0.031 & 0.199 & -0.201* & 0.367*** \\
        & (0.112) & (0.110) & (0.109) & (0.110) & (0.091) & (0.107) \\
        Switzerland & -0.165 & -0.106 & 0.090 & 0.079 & -0.264* & 0.069 \\
        & (0.135) & (0.129) & (0.140) & (0.138) & (0.124) & (0.134) \\
        France & 0.459*** & 0.528*** & 0.655*** & 0.546*** & -1.117*** & 0.723*** \\
        & (0.110) & (0.117) & (0.111) & (0.114) & (0.121) & (0.116) \\
        Poland & 0.038 & 0.346* & 0.547*** & 0.141 & -0.429** & 0.423** \\
        & (0.141) & (0.154) & (0.146) & (0.149) & (0.159) & (0.155) \\
        Sweden & 0.684* & 0.553 & -0.139 & 0.762** & -1.020*** & 0.601* \\
        & (0.272) & (0.408) & (0.285) & (0.259) & (0.276) & (0.305) \\
        Belgium & -0.113 & -0.394 & -0.404 & -0.247 & -0.139 & 0.000 \\
        & (0.245) & (0.279) & (0.255) & (0.270) & (0.208) & (0.247) \\
        Netherlands & -0.145 & 0.211 & 0.023 & -0.091 & 0.040 & 0.209 \\
        & (0.170) & (0.181) & (0.183) & (0.154) & (0.149) & (0.178) \\
        Spain & -0.148 & -0.158 & 0.464*** & -0.117 & -0.575*** & 0.492*** \\
        & (0.090) & (0.096) & (0.093) & (0.090) & (0.091) & (0.094) \\
        Portugal & 0.554*** & 0.859*** & 1.382*** & 0.828*** & -1.079*** & 1.240*** \\
        & (0.138) & (0.141) & (0.134) & (0.136) & (0.141) & (0.141) \\
        Italy & 0.132 & 0.265 & 0.869*** & 0.173 & -1.038*** & 0.470* \\
        & (0.170) & (0.186) & (0.192) & (0.172) & (0.174) & (0.192) \\
        Other & -0.033 & 0.463 & 1.097*** & 0.439 & -1.218*** & 0.722* \\
        & (0.309) & (0.306) & (0.256) & (0.347) & (0.271) & (0.364) \\
        \bottomrule
        Observations & 3492 & 3492 & 3492 & 3492 & 3492 & 3492 \\
		\bottomrule
	\end{tabular}
        \small
    \raggedright
    \break
    \textit{Note.} All models include country fixed effects (Germany is the baseline). Ordered logit coefficients are reported. Robust standard errors in parentheses. * \textit{p} < .05, ** \textit{p} < .01, *** \textit{p} < .001. \\
	\label{tab:full_country}
\end{table}

\begin{table}[ht]
	\caption{Predicting level of stigma using assistant type, age, gender, and country of residence via ordered logistic regression. Model (1) includes interaction between treatment and age, Model (2) includes interaction between treatment and gender.}
	\centering
	\begin{tabular}{lcc}
		\toprule
		   & Model (1): Interaction Age & Model (2): Interaction Gender \\
		\midrule
        Assistant type (Human = 1, AI = 0): & 0.605*** & 0.569*** \\
         & (0.060) & (0.081) \\
        Age (centered) & -0.008* &  \\
         & (0.004) &  \\
        Human x Age (centered) & 0.013* &  \\
         & (0.005) &  \\
        Gender female & -0.193** & -0.234** \\
         & (0.060) & (0.086) \\
        Age &  & -0.002 \\
         &  & (0.003) \\
        Human x Gender &  & 0.083 \\
         &  & (0.120) \\
        \bottomrule
        Observations & 3492 & 3492 \\
        \bottomrule
	\end{tabular}
        \small
    \raggedright
    \break
    \textit{Note.} All models include country fixed effects (not shown for clarity). Age is centered when included in the interaction term. Ordered logit coefficients are reported. Robust standard errors in parentheses. * \textit{p} < .05, ** \textit{p} < .01, *** \textit{p} < .001. \\
	\label{tab:interaction_preferences}
\end{table}

\begin{table}[ht]
	\caption{ Descriptive statistics for social preferences and emotions across countries.}
	\centering
	\begin{tabular}{lccccccc}
		\toprule
		   & Fairness & Trust & Reciprocity & Efficiency & Stigma & Empathy & n \\
		\midrule
        Germany & 4.26 (0.78) & 4.19 (0.90) & 3.73 (1.07) & 4.28 (0.76) & 2.45 (0.91) & 3.89 (0.88) & 1036 \\
        Austria & 4.37 (0.73) & 4.26 (0.85) & 3.69 (1.11) & 4.34 (0.75) & 2.35 (0.86) & 4.03 (0.90) & 412 \\
        Switzerland & 4.17 (0.80) & 4.15 (0.81) & 3.72 (1.11) & 4.29 (0.77) & 2.31 (0.90) & 3.88 (0.92) & 205 \\
        France & 4.40 (0.76) & 4.37 (0.86) & 4.03 (1.06) & 4.43 (0.77) & 1.98 (1.09) & 4.14 (1.00) & 421 \\
        Poland & 4.26 (0.75) & 4.31 (0.81) & 3.99 (0.91) & 4.32 (0.75) & 2.24 (1.01) & 4.02 (0.95) & 183 \\
        Netherlands & 4.20 (0.79) & 4.31 (0.76) & 3.75 (1.01) & 4.30 (0.61) & 2.47 (0.84) & 4.00 (0.81) & 101 \\
        Spain & 4.20 (0.79) & 4.10 (0.95) & 3.98 (1.01) & 4.24 (0.77) & 2.17 (1.02) & 4.10 (0.92) & 647 \\
        Portugal & 4.39 (0.86) & 4.48 (0.83) & 4.36 (0.92) & 4.53 (0.73) & 1.97 (1.04) & 4.35 (0.90) & 293 \\
        Italy & 4.35 (0.65) & 4.33 (0.84) & 4.20 (0.96) & 4.37 (0.67) & 1.93 (0.92) & 4.11 (0.89) & 101 \\
		\bottomrule
	\end{tabular}
        \small
    \raggedright
    \break
    \textit{Note.}Mean values are reported and standard deviations in parentheses. Ratings were indicated on a scale from 1 (strongly disagree) to 5 (strongly agree). \\
	\label{tab:descriptives}
\end{table}

\begin{table}[ht]
	\caption{Predicting social preference rating using assistant type, age, gender, and country of residence via ordered logistic regression, including interaction terms between assistant type and country. }
	\centering
	\begin{tabular}{lcccccc}
		\toprule
		   & Fairness & Trust & Reciprocity & Efficiency & Stigma & Empathy \\
		\midrule
        Assistant type (Human = 1, AI = 0) & 0.365** & -0.098 & 0.015 & -0.104 & 0.747*** & 0.272* \\
        & (0.113) & (0.118) & (0.113) & (0.112) & (0.107) & (0.113) \\
        
        Germany & 0.000 & 0.000 & 0.000 & 0.000 & 0.000 & 0.000 \\
        & (.) & (.) & (.) & (.) & (.) & (.) \\
        
        Austria & 0.306 & 0.101 & -0.168 & 0.228 & -0.115 & 0.298 \\
        & (0.158) & (0.160) & (0.155) & (0.159) & (0.148) & (0.159) \\
        
        Switzerland & -0.148 & -0.163 & -0.106 & 0.175 & -0.032 & -0.009 \\
        & (0.192) & (0.199) & (0.200) & (0.199) & (0.187) & (0.196) \\
        
        France & 0.613*** & 0.657*** & 0.660*** & 0.865*** & -1.132*** & 0.985*** \\
        & (0.152) & (0.166) & (0.156) & (0.162) & (0.159) & (0.162) \\
        
        Poland & -0.134 & 0.033 & 0.367 & -0.124 & -0.166 & 0.112 \\
        & (0.193) & (0.204) & (0.194) & (0.194) & (0.198) & (0.202) \\
        
        Netherlands & -0.006 & 0.227 & 0.064 & 0.058 & 0.231 & 0.081 \\
        & (0.255) & (0.277) & (0.276) & (0.258) & (0.245) & (0.269) \\
        
        Spain & -0.131 & -0.386** & 0.263* & -0.163 & -0.434*** & 0.537*** \\
        & (0.128) & (0.134) & (0.130) & (0.128) & (0.126) & (0.133) \\
        
        Portugal & 0.774*** & 1.062*** & 1.402*** & 1.207*** & -1.125*** & 1.532*** \\
        & (0.178) & (0.197) & (0.185) & (0.193) & (0.181) & (0.190) \\
        
        Italy & 0.336 & 0.204 & 1.272*** & 0.584* & -1.266*** & 0.685* \\
        & (0.178) & (0.289) & (0.293) & (0.280) & (0.272) & (0.281) \\
        
        Assistant type x Germany & 0.000 & 0.000 & 0.000 & 0.000 & 0.000 & 0.000 \\
        & (.) & (.) & (.) & (.) & (.) & (.) \\
        
        Assistant type x Austria & 0.068 & 0.249 & 0.383 & -0.059 & -0.167 & 0.124 \\
        & (0.216) & (0.221) & (0.213) & (0.214) & (0.200) & (0.217) \\
        
        Assistant type x Switzerland & -0.036 & 0.124 & 0.387 & -0.187 & -0.460 & 0.159 \\
        & (0.273) & (0.281) & (0.279) & (0.277) & (0.264) & (0.281) \\
        
        Assistant type x France & -0.333 & -0.200 & 0.026 & -0.584** & 0.005 & -0.493* \\
        & (0.213) & (0.227) & (0.215) & (0.219) & (0.214) & (0.220) \\
        
        Assistant type x Poland & 0.386 & 0.682* & 0.338 & 0.598* & -0.554 & 0.656* \\
        & (0.288) & (0.309) & (0.293) & (0.293) & (0.286) & (0.302) \\
        
        Assistant type x Netherlands & -0.285 & -0.016 & -0.063 & -0.290 & -0.385 & 0.255 \\
        & (0.364) & (0.387) & (0.377) & (0.358) & (0.349) & (0.380) \\
        
        Assistant type x Spain & -0.131 & -0.386** & 0.263* & -0.163 & -0.326 & -0.071 \\
        & (0.128) & (0.134) & (0.130) & (0.128) & (0.177) & (0.187) \\
        
        Assistant type x Portugal & 0.774*** & 1.062*** & 1.402*** & 1.207*** & 0.075 & -0.525* \\
        & (0.178) & (0.197) & (0.185) & (0.193) & (0.247) & (0.262) \\
        
        Assistant type x Italy & 0.336 & 0.204 & 1.272*** & 0.584* & 0.386 & -0.369 \\
        & (0.258) & (0.289) & (0.293) & (0.280) & (0.372) & (0.392) \\
        
        Age & 0.007* & 0.012*** & 0.015*** & 0.006* & -0.002 & 0.008** \\
        & (0.003) & (0.003) & (0.003) & (0.003) & (0.003) & (0.003) \\
        
        Gender female & 0.366*** & 0.401*** & 0.504*** & 0.407*** & -0.176** & 0.487*** \\
        & (0.063) & (0.066) & (0.064) & (0.064) & 0.747*** & 0.272* \\
        \bottomrule
        Observations & 3380 & 3380 & 3380 & 3380 & 3380 & 3380 \\
        \bottomrule
	\end{tabular}
        \small
        \raggedright
        \break
        \textit{Note.}All models include country fixed effects (Germany is baseline). Ordered logit coefficients are reported. Robust standard errors in parentheses. * \textit{p} < .05, ** \textit{p} < .01, *** \textit{p} < .001. \\
	\label{tab:interaction_country}
\end{table}

\end{document}